\documentclass[aps,reprint,nofootinbib]{revtex4-2}
\usepackage{amsmath,amssymb}
\usepackage{hyperref}
\usepackage{color}
\usepackage{graphicx}

\newcommand{\RE}{\operatorname{Re}}
\newcommand{\tr}{\operatorname{tr}}

\newcommand{\Neg}{\mathcal{N}}

\begin{document}
\title{Entanglement oscillations from flavour coherence}

\author{J. A.~Aguilar-Saavedra}
\affiliation{Instituto de F\'isica Te\'orica IFT-UAM/CSIC, c/Nicol\'as Cabrera 13--15, 28049 Madrid, Spain} 
\author{I. Bachiller Perea }
\affiliation{Institute of Physics, Ecole Polytechnique F\'ed\'erale de Lausanne (EPFL), Lausanne, Switzerland} 

\begin{abstract}
We investigate the oscillating behaviour of the spin entanglement generated by flavour coherence in $B$-meson decays, as a function of the proper decay time. In $B_s^0\to J/\psi\,\phi$, its time dependence can be certified even without full spin tomography, with a projected sensitivity exceeding $4\sigma$ with current LHCb data. For $B_s^0\to\phi\gamma$, photon conversion polarimetry offers prospects for evidence of entanglement oscillations at future $Z$ factories.
\end{abstract}

\maketitle

\section{Introduction}

Quantum entanglement~\cite{Schrodinger1935} is one of the features of quantum mechanics that most clearly distinguish it from classical physics. Physical interactions can generate entanglement, modify its amount or transfer it between degrees of freedom. High-energy physics experiments allow us to test quantum-information concepts at the highest energy scales. Tests of spin entanglement have been proposed for top quarks~\cite{Afik:2020onf,Severi:2021cnj,Afik:2022kwm,Aguilar-Saavedra:2022uye, Dong:2023xiw,Han:2023fci,Aguilar-Saavedra:2023hss,Aguilar-Saavedra:2024hwd,Maltoni:2024csn,Cheng:2024btk,Aoude:2025ovu,Aoude:2026eeg,Belmonte:2026yno}, $b$-quarks~\cite{Afik:2025grr}, leptons~\cite{Altakach:2022ywa,Aguilar-Saavedra:2023lwb,Ehataht:2023zzt,Fabbrichesi:2024wcd,Han:2025ewp,Zhang:2025mmm} and weak bosons~\cite{Barr:2021zcp,Aguilar-Saavedra:2022wam,Ashby-Pickering:2022umy,Aguilar-Saavedra:2022mpg,Fabbrichesi:2023cev,Morales:2023gow,Bernal:2024xhm,Ruzi:2024cbt,Wu:2024ovc, Bernal:2025zqq}. To date, measurements have been performed in top-quark pair production~\cite{ATLAS:2023fsd,CMS:2024pts,CMS:2024zkc} and Higgs decays into $Z$-boson pairs~\cite{ATLAS:2026hye} at the Large Hadron Collider (LHC)

Neutral $B$ mesons provide a natural setting in which to address the interplay between flavour evolution, decay and entanglement. In the $B_d^0$--$\bar B_d^0$ and $B_s^0$--$\bar B_s^0$ systems, the flavour eigenstates are superpositions of two mass eigenstates. Their coherent propagation gives rise to particle--antiparticle oscillations, with frequencies determined by the corresponding mass splittings~\cite{PDB}. A meson produced with a definite flavour evolves into a coherent superposition of flavour states. Therefore, if both flavour states can decay into the same final state, interference between their decay amplitudes can make the spin state of the decay products depend on the meson decay time.

Here we show that flavour coherence can induce an oscillatory dependence of the spin entanglement of the decay products. The relevant time is the proper decay time of the parent meson: events in different time intervals prepare different final-state spin density matrices. (Once the decay takes place, the spin entanglement is fixed.) Oscillations in the amount of entanglement have previously been observed in photonic simulations of non-Markovian dynamics~\cite{Cialdi:2010tsd} and in a superconducting qubit coupled to a microscopic two-level system~\cite{Sun:2012wqx}.

We study this mechanism in $B_s^0 \to J/\psi \,\phi$ and $B_s^0 \to \phi \gamma$ decays tagged by the initial flavour. For $B_s^0\to J/\psi\,\phi$, we find that the large samples collected at LHC Runs 1--3 allow for strong evidence of the time dependence of the entanglement beyond the $4\sigma$ level. For $B_s^0 \to \phi \gamma$, the required statistics to have evidence of the effect could be achieved at future $Z$ factories.

\section{Spin entanglement from flavour coherence}

The flavour state of a neutral $B$ meson can be described in the basis $\{ |B^0\rangle, |\bar B^0\rangle \}$ by an unnormalised density operator
\begin{equation}
\tilde \rho_B(t) = \frac{1}{2} \left[ \tr \tilde \rho_B(t) \,\openone_2 + \sum_{i=1}^3 \tilde b_i(t) \sigma_i \right] \,,
\end{equation}
where $\sigma_i$, $i=1,2,3$, are the Pauli matrices and the coefficients $\tilde b_i(t)$ are real. Provided the state at $t=0$ is normalised, $\tr \tilde\rho_B(t)$ gives the survival probability up to time $t$. The normalised operator
\begin{equation}
\rho_B(t) = \frac{1}{2} \left[ \openone_2 + \sum_{i=1}^3 b_i(t) \sigma_i \right] 
\end{equation}
specifies the flavour state at a time $t$ conditional on survival. The normalised quantities $b_i(t) = \tilde b_i(t) / \tr \tilde\rho_B(t)$ are the components of its Bloch vector. The flavour populations are $(1 \pm b_3(t))/2$, and $(\rho_B)_{12} = (b_1(t) - i b_2(t))/2$ describes their coherence.  The state of the meson can thereby be regarded as a qubit in flavour space, in full analogy with the spin state of a spin-$1/2$ particle.

Neglecting CP violation in mixing (which is an excellent approximation~\cite{PDB}), we can choose the phases such that the mass eigenstates are
\begin{equation}
|B_{L,H}\rangle =\frac{|B^0\rangle \pm |\bar B^0\rangle}{\sqrt 2} \,.
\label{ec:BLH}
\end{equation}
We denote the mass and width of the heavy ($H$) and light ($L$) eigenstates as $m_{H,L}$, $\Gamma_{H,L}$, and define $\Gamma = (\Gamma_L + \Gamma_H)/2$, $\Delta\Gamma = \Gamma_L - \Gamma_H$ and $\Delta m = m_H - m_L$. Then, the time evolution is~\cite{Cheng:2025zcf}
\begin{eqnarray}
\tr \tilde \rho_B(t) & = & e^{-\Gamma t}\left[ \cosh (\Delta\Gamma t/2) -b_1^0 \sinh (\Delta \Gamma t/2) \right] \,, \notag \\
\tilde b_1(t) & = & e^{-\Gamma t} \left[ b_1^0 \cosh (\Delta\Gamma t/2) 
- \sinh (\Delta\Gamma t/2) \right],\notag \\
\tilde b_2(t) & = & e^{-\Gamma t} \left[ b_2^0 \cos(\Delta m t) + b_3^0 \sin(\Delta m t)
\right] \,, \notag \\
\tilde b_3(t) & = &e^{-\Gamma t} \left[ b_3^0 \cos(\Delta m t) - b_2^0 \sin(\Delta m t)
\right] \,,
\end{eqnarray}
where $b_i^0=\tilde b_i(0)$ specify the initial flavour state. A pure $B^0$ state at $t=0$ is characterised by $\vec b^0 = (0,0,1)$, while a $\bar B^0$ state corresponds to $\vec b^0 = (0,0,-1)$.

After the meson decay, the spin state of its decay products can be described using the formalism of post-decay density operators~\cite{Aguilar-Saavedra:2024fig}. For a given decay channel and fixed final-state momenta $P$, let $|\xi_\alpha\rangle$ denote a basis of the product spin space of the decay
products, and $M_{\alpha i} = \langle P \xi_\alpha | T | i \rangle$ the decay amplitudes for a meson state $|i\rangle$, with $T$ the transition operator. The matrix $M$ has two columns $M_{\alpha 1} = A_\alpha$ and $M_{\alpha 2} = \bar A_\alpha$, corresponding to the $B^0$ and $\bar B^0$ decay amplitudes, respectively.
In the convention of Eq.~(\ref{ec:BLH}), the barred amplitudes carry the relative weak phase between direct decay and decay after mixing.  The unnormalised spin density operator is
\begin{equation}
\tilde \rho(t) = \sum_{\alpha \beta} (M \rho_B(t) M^\dagger)_{\alpha \beta} |\xi_\alpha \rangle \langle \xi_\beta| \,.
\label{ec:rhoSt}
\end{equation}
The off-diagonal entries of $\rho_B(t)$ generate interference between the two flavour decay amplitudes, while its diagonal entries weight their separate contributions.
Thus, the spin entanglement between the decay products can depend on the proper decay time of the parent meson. The normalised spin density operator is
\begin{equation}
\rho(t) = \frac{\tilde \rho(t)}{\tr \tilde \rho(t)} \,.
\label{ec:rho}
\end{equation}
At fixed momenta and decay time, a pure flavour state gives a pure spin state. For a finite phase-space region the operator (\ref{ec:rhoSt}) must be integrated over that region before normalisation. Likewise, for a finite decay-time interval, the spin density operator is obtained by integrating $M\tilde\rho_B(t)M^\dagger$ over the interval before normalisation.

\section{Entanglement oscillations in $B_s^0$ decays}

We now particularise for $B_s^0$ mesons, whose oscillation period is much shorter than their lifetime, allowing several oscillations to occur before most mesons decay. We use as numerical inputs $\Gamma_s = 0.66$ ps$^{-1}$, $\Delta \Gamma_s = 0.0783$ ps$^{-1}$ and $\Delta m_s = 17.766$ ps$^{-1}$~\cite{HeavyFlavorAveragingGroupHFLAV:2024ctg}. The density operator $\rho$ in (\ref{ec:rho}) describes the spin state of the two decay products. We quantify spin entanglement using the negativity~\cite{Vidal:2002zz},
\begin{equation}
\Neg(\rho)=\frac{\|\rho^{T_2}\|_1-1}{2}\,,
\end{equation}
where $T_2$ denotes partial transposition with respect to the second subsystem (the spin space of the second particle) and $\|X\|_1=\tr\sqrt{X^\dagger X}$ is the trace norm. (The negativity is the same when taking partial transpose on the first subsystem.) The negativity is the sum of the moduli of the negative eigenvalues of $\rho^{T_2}$. By the Peres-Horodecki criterion~\cite{Peres:1996dw,Horodecki:1996nc}, $\Neg>0$ is a sufficient condition for entanglement. 

We quantise both spins along a common axis, chosen along the $J/\psi$ momentum for $B_s^0\to J/\psi \,\phi$ and along the photon momentum for $B_s^0 \to \phi \gamma$, taken in the $B_s^0$ meson rest frame. We use the single-particle spin ordering $(+,0,-)$, with the
zero eigenvalue omitted for the photon. In the $J/\psi \,\phi$ channel, angular momentum conservation restricts the allowed spin states to
$\{ |+-\rangle, |00\rangle, |-+\rangle\}$, a subset of the nine states in the full product basis,
with the $J/\psi$ spin listed first. For $\phi\gamma$, the allowed states are
$\{ |+-\rangle, |-+\rangle\}$, with the photon spin listed first.
Labelling the matrix elements of the spin density operator as $\rho_{m_1 m_2;n_1n_2} \equiv \langle m_1 m_2| \rho |n_1 n_2 \rangle$, the corresponding negativities are
\begin{align}
& \Neg_{J/\psi\phi} = |\rho_{+-;00}| + |\rho_{00;-+}| + |\rho_{+-;-+}| \,, \notag \\
& \Neg_{\phi\gamma} = |\rho_{+-;-+}| \,,
\label{ec:neg}
\end{align}
where we omit the time dependence for brevity.
The structure of the density operator imposed by angular momentum conservation makes $\Neg > 0$ also a necessary condition for entanglement in both decay modes~\cite{Aguilar-Saavedra:2022wam}. The maximal values are $\Neg_{J/\psi\phi}=1 $ and $\Neg_{\phi\gamma}=1/2$, attained by maximally-entangled states in the respective product spaces of dimensions $3 \times 3$ and $3 \times 2$.

The relation between flavour coherence and spin entanglement is particularly simple for
$B_s^0 \to \phi\gamma$ in the CP-conserving limit, keeping only the dominant photon helicity for each flavour. In this limit, $A_\alpha = (1,0)^T$ and $\bar A_\alpha = (0,1)^T$, up to normalisation. The two flavours then decay into orthogonal product spin states, $|B_s^0 \rangle \to |+-\rangle$, $|\bar B_s^0 \rangle \to |-+\rangle$, and the negativity directly measures the flavour coherence,
\begin{equation}
\Neg_{\phi\gamma}(t) = \left| (\rho_B(t))_{12} \right| =\frac{1}{2} \left[ b_1(t)^2 + b_2(t)^2 \right]^{1/2} \,.
\end{equation}
For an initially pure $B_s^0$ or $\bar B_s^0$ state, this gives
\begin{equation}
\Neg_{\phi\gamma}(t) = \frac{\left[ \sinh^2(\Delta\Gamma_s t/2) + \sin^2(\Delta m_s t) \right]^{1/2} }{2\cosh(\Delta\Gamma_s t/2)} \,.
\end{equation}
Neglecting the width difference, this reduces to $\Neg_{\phi\gamma}(t)=|\sin(\Delta m_s t)|/2$.
The negativity oscillates with period $T_s/2$, where
$T_s=2\pi/\Delta m_s\simeq 0.354~\mathrm{ps}$ is the flavour-oscillation
period. It vanishes when the meson has a definite flavour at decay and reaches $1/2$ when the two flavours are equally probable.
We have checked the dominant-helicity approximation using the QCD-factorisation calculation of Ref.~\cite{Ball:2006eu} with inputs from Refs.~\cite{Ball:2007rt,Ball:2007zt}. The two dominant helicity amplitudes have nearly equal magnitudes, and in each case the opposite-helicity amplitude is about $1\%$ as large. The resulting corrections to the negativity oscillations are small.

For $B_s^0\to J/\psi\phi$, we obtain the amplitudes from the polarisation fractions and relative strong phases measured in the LHCb time-dependent angular analysis~\cite{LHCb:2023sim}, translating the transversity amplitudes to our spin basis using the angular conventions of Ref.~\cite{LHCb:2019nin}. Up to a common normalisation, this gives
\begin{align}
& A_{+-}=-0.684342 + 0.081422i \, \notag \\
& A_{00}=0.719650 \,, \notag \\
& A_{-+}=-0.002379 - 0.084449i \,.
\end{align}
Neglecting direct CP violation and subleading weak phases in the decay amplitudes, the corresponding $\bar B_s^0$ amplitudes in our flavour-phase convention are $\bar A_{+-} = e^{-i\phi_s}A_{-+}$, $\bar A_{00} = e^{-i\phi_s}A_{00}$, $\bar A_{-+} = e^{-i\phi_s} A_{+-}$. The phase $\phi_s = -0.0377$~\cite{Charles:2015gya} describes CP violation in the interference between mixing and decay.

\begin{figure}[htb]
\begin{center}
\includegraphics[width=8cm,clip=]{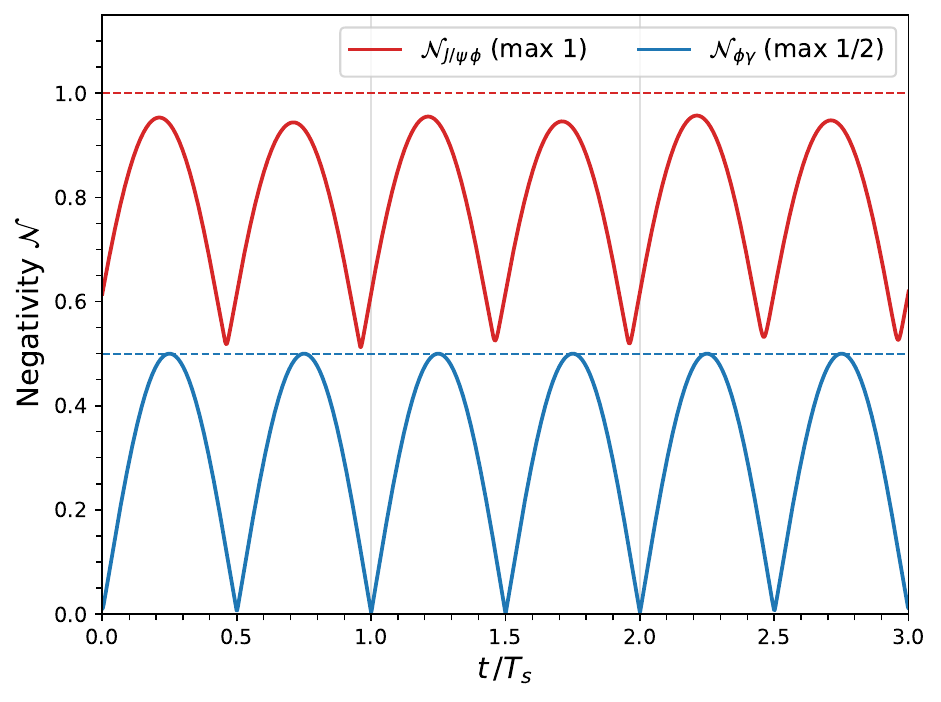}
\caption{Negativity of the $J/\psi \,\phi$ (red) and $\phi\gamma$ (blue) spin states from an initially pure $B_s^0$. Dashed lines indicate the maximal values.}
\label{fig:NTH}
\end{center}
\end{figure}

For an initial $B_s^0$ state, the resulting negativities are shown in Fig.~\ref{fig:NTH} as functions of $t/T_s$. Both channels exhibit pronounced oscillations: the $\phi \gamma$ spin state evolves between nearly separable and maximally entangled configurations, while the $J/\psi\, \phi$ spin state remains strongly entangled throughout the displayed interval.

\section{$B_s^0\to J/\psi\,\phi$ analysis}

For this decay channel we consider the decays $J/\psi\to\mu^+\mu^-$ and $\phi\to K^+K^-$, which allow us to measure matrix elements of the $J/\psi \, \phi$ spin density operator. We label as $\Omega_\mu=(\theta_\mu,\varphi_\mu)$, $\Omega_K=(\theta_K,\varphi_K)$ the polar and azimuthal angles, respectively, of $\mu^-$ and $K^-$ in the rest frame of their parent 
particles. The differential decay distribution is
\begin{equation}
\frac{1}{\Gamma_f} \frac{d\Gamma_f}{d\Omega_\mu d\Omega_K} = \frac{9}{16\pi^2} \sum \rho_{i_1 i_2 ; j_1 j_2} \Gamma^{J/\psi}_{i_1 j_1}(\Omega_\mu) \Gamma^\phi_{i_2 j_2} (\Omega_K) \,.
\end{equation}
Here $\Gamma_f$ denotes the partial decay width into the final
state $f=J/\psi\, \phi$, with $d\Omega = d\!\cos\theta d\phi$. The decay matrices are $\Gamma^{J/\psi}(\Omega) = 
\left[\openone_3-v(\Omega)v^\dagger(\Omega)\right]/2$,  $\Gamma^\phi(\Omega) = v(\Omega) v^\dagger(\Omega)$~\cite{Rahaman:2021fcz}, with 
\begin{equation}
v(\Omega) = \left( -\frac{\sin\theta}{\sqrt{2}}e^{i\varphi}, \cos\theta, \frac{\sin\theta}{\sqrt{2}}e^{-i\varphi} \right)^T \,.
\end{equation}
Let us define for brevity $q=\rho_{00;00}$, $w=\rho_{+-;00}+\rho_{00;-+}$,
$z=\rho_{+-;-+}$, $c_a=\cos\theta_a$, $s_a=\sin\theta_a$, $\varphi=\varphi_\mu-\varphi_K$. The differential distribution reads
\begin{eqnarray}
\frac{1}{\Gamma_f} \frac{d\Gamma_f}{dc_\mu\,dc_K\,d\varphi} & = &
\frac{9}{64\pi} \left[ (1-q)(1+c_\mu^2)s_K^2 +4q\,s_\mu^2c_K^2 \right. \notag \\
& & +\sin2\theta_\mu\sin2\theta_K \RE (we^{i\varphi}) \notag \\
& & \left. -2s_\mu^2s_K^2 \RE(ze^{2i\varphi}) \right] \,.
\end{eqnarray}
We point out that the two off-diagonal elements $\rho_{+-;00}$, $\rho_{00;-+}$ entering the negativity (\ref{ec:neg}) cannot be determined separately from this distribution.
The reason can be traced back to the vector interaction mediating $J/\psi \to \mu^+\mu^-$, which makes the angular distribution insensitive to the vector polarisation of the $J/\psi$, while retaining sensitivity to its tensor polarisation.
Nevertheless, positivity of $\rho$ implies the sharp bounds $L \leq \mathcal{N}_{J/\psi\phi} \leq U$, with
\begin{equation}
L = |z|+|w| \,,\quad U = |z|+\sqrt{q(1-q+2|z|)} \,.
\label{ec:bounds}
\end{equation}
The lower bound trivially results from the triangle inequality, while the upper bound can be obtained using the Schwarz inequality and the fact that $\rho$ is positive semidefinite. If the lower bound in one decay-time bin exceeds the upper bound in another, the two bins must contain spin states with different negativities.

In the CP-conserving limit, oscillations from initially pure $B_s^0$ and $\bar B_s^0$ states give the same $q$ and conjugate values of $w$ and $z$. We combine the samples by reversing $\varphi$ for the $\bar B_s^0$ tag. For equal production rates and charge-symmetric selection, the terms linear in $\phi_s$ cancel, leaving a tiny correction of order $\phi_s^2$.

The rapid flavour oscillations make time resolution essential. We use a Gaussian decay-time resolution $\sigma_t=0.042$ ps~\cite{LHCb:2023sim}. As previously discussed, initial flavour tagging is also necessary. 
We consider a tighter selection than the nominal one in Ref.~\cite{LHCb:2025okz}. From the simulated distributions in Fig.~2 of that reference and the corresponding data calibration, we estimate a retained signal fraction $\epsilon\simeq0.04$ and an average mistag probability $\bar\omega\simeq0.15$.
The same approximate benchmark is obtained from a study of the CMS mistag distributions~\cite{CMS:2024znt}. A small mistag probability is essential since mistagging mixes the two production flavours and reduces the flavour-oscillation terms by $1-2\bar\omega$.

\begin{figure}[htb]
\begin{center}
\includegraphics[width=8cm,clip=]{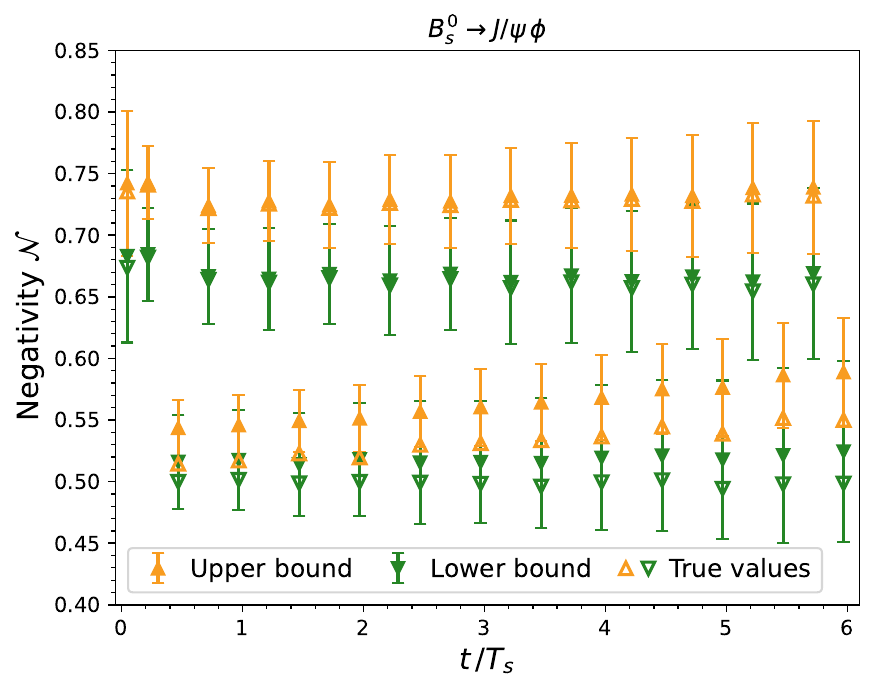}
\caption{Expected bounds on the $J/\psi\,\phi$ negativity for $8\times10^4$ tagged decays. Filled symbols represent the Monte Carlo estimates $(\widehat L_i,\widehat U_i)$, with $1\sigma$ statistical uncertainties, while open symbols indicate the true values $(L_i, U_i)$.}
\label{fig:NJpsi}
\end{center}
\end{figure}

We estimate the sensitivity using 1000 pseudoexperiments with $8 \times 10^4$ events each, including the aforementioned mistag probability and time resolution. This sample size corresponds to the selection efficiency $\epsilon \simeq 0.04$ applied to an assumed total of $2 \times 10^6$ reconstructed signal decays from the LHCb Run~1--2 datasets (which include the signal losses from background rejection) together with the expected Run~3 yield, obtained by rescaling the Run~2 sample to the
$26.7$ fb$^{-1}$ recorded in 2024--2026~\cite{Alessio:2026zzz}. We assume uniform angular and decay-time acceptance. The reconstructed times are divided into bins of nominal width $T_s/4$, with centres chosen near the predicted maxima and minima, except for the first bin. Within each bin $i$, $q_i$, $w_i$ and $z_i$ are determined from the angular distributions, providing estimates $\widehat L_i$, $\widehat U_i$ of the negativity bounds $L_i$, $U_i$. Figure~\ref{fig:NJpsi} shows the mean bounds and their standard deviations over the pseudoexperiments. The upward shifts in $\widehat L_i$, $\widehat U_i$ with respect to the true values are caused by statistical fluctuations in the estimates of $q_i$, $w_i$ and $z_i$, which enter the bounds in Eq.~(\ref{ec:bounds}) nonlinearly. The oscillatory behaviour of the negativity is apparent, but its quantification in terms of a $p$-value is more delicate.

If the negativity is constant, the average lower bound at the peaks cannot exceed the average upper bound at the troughs. To test this ordering, instead of directly using (\ref{ec:bounds}) we use 
\begin{equation}
L'_i = \RE\!\left(e^{-i\alpha_i}w_i+e^{-i\beta_i}z_i\right) \,, \quad
U'_i = \frac{1}{2} +2|z_i| \,,
\label{ec:bounds2}
\end{equation}
where $L'_i\leq L_i$ for any phases $\alpha_i,\beta_i$, and $U_i\leq U'_i$. A subset of `reference' bins is used to determine the phases and the peak and trough assignments, which are considered fixed thereafter.
Those reference bins are not used for the subsequent $p$-value calculation. The choice of bounds (\ref{ec:bounds2}) improves the sensitivity. For lower bounds at the peaks, when $w_i=0$, $\RE(e^{-i\alpha_i}\widehat w_i)$ averages to zero over the different pseudoexperiments, while $|\widehat w_i|$ has a positive mean (and likewise when $z_i = 0$). For the upper bounds at the troughs, $U'_i$ is close to $U_i$ and avoids fluctuations of $\widehat q_i$ under the square root. We use the test statistic
\begin{equation}
T = \frac{\sum_{i \in \mathcal{P}} n_i\widehat L'_i}{\sum_{i\in \mathcal{P}} n_i}
  - \frac{\sum_{i\in \mathcal{T}} n_i\widehat U'_i}{\sum_{i\in \mathcal{T}} n_i} \,,
\label{ec:TJpsi}
\end{equation}
where $\mathcal{P}$ and $\mathcal{T}$ denote the peak and trough bins, $n_i$ is the number of
events in bin $i$, and the hats denote bounds evaluated with the measured
$\hat w_i$ and $\hat z_i$. Large positive values of $T$ disfavour the null hypothesis of constant negativity. For $8\times10^4$ tagged decays, the (conservative) test gives an expected $4.4\sigma$ sensitivity for rejecting constant negativity. The sensitivities for $6\times10^4$ and $10^5$ events, corresponding to $\epsilon = 0.03,0.05$, are $3.4\sigma$ and $5.2\sigma$, respectively. Further details on the statistical test are given in the Appendix.

\section{$B_s^0\to\phi\gamma$ analysis}

In this process the photon polarisation can be accessed when photons undergo nuclear
Bethe--Heitler conversion into an $e^+e^-$ pair~\cite{Bishara:2013vya,Bishara:2015yta}. 
The formalism is similar to the previous decay channel, but replacing the $J/\psi$ decay matrix $\Gamma^{J/\psi}_{i_1 j_1}(\Omega_\mu)$ by a photon conversion matrix
\begin{equation}
H_{\lambda\lambda'}(\xi) =\sum_s\mathcal A_{\lambda s}(\xi) \mathcal A_{\lambda's}^{*}(\xi) \,, 
\end{equation}
where $\mathcal A_{\lambda s}$ are the conversion amplitudes for photon helicity $\lambda$, with unobserved $e^+ e^-$ spins labelled by $s$, and $\xi$ denote the conversion variables, i.e. the lepton energies and angles in the laboratory frame. Denoting by 
$d\mu(\xi)$ the phase-space measure for the conversion variables, the joint distribution is
\begin{equation}
\frac{d\Gamma_f}{d\Omega_K\,d\mu(\xi)} \propto \sum_{i_1i_2j_1j_2} \rho_{i_1i_2;j_1j_2}\,
H_{i_1j_1}(\xi)\, \Gamma^\phi_{i_2j_2}(\Omega_K) \,.
\label{ec:rhoGG2}
\end{equation}
For our sensitivity estimates we perform a partial integration
over the conversion kinematics. Let us define $\chi=(\varphi_+ + \varphi_-)/2$, where $\varphi_\pm$ are the lepton azimuthal angles around the photon direction, and $\eta$ the rest of conversion variables. For unpolarised material at Born level the two diagonal elements of $H$ are equal. Furthermore, rotational invariance around the photon direction allows to write
$H_{++}=H_{--} \equiv H_0(\eta)$, $H_{+-}=H_{-+}^*=H_0(\eta)h(\eta)e^{2i\chi}$,
with $|h(\eta)|\leq1$ from positivity. For a $\chi$-independent selection,
$d\mu(\xi) = d\chi\,d\mu(\eta)$, and integration over $\eta$
gives an effective matrix which, upon normalisation, reads
\begin{equation}
H(\chi,\eta) = \frac{1}{2}
\left( \begin{array}{cc}
1 & \kappa e^{2i\chi} \\
\kappa^* e^{-2i\chi} & 1
\end{array} \right) \,,
\label{ec:Hfinal}
\end{equation}
where
\begin{equation}
\kappa= \displaystyle\int d\mu(\eta)\,H_0(\eta)h(\eta) \left/ \displaystyle\int d\mu(\eta)\,H_0(\eta) \right. \,.
\label{ec:kint}
\end{equation}
If the selection is symmetric under $\varphi_\pm\to-\varphi_\pm$,
$\kappa$ is real. Plugging (\ref{ec:Hfinal}) and $\Gamma^\phi$ into (\ref{ec:rhoGG2}) and writing $\varphi \equiv \chi-\varphi_K$, $z \equiv \rho_{+-;-+}$, the normalised distribution is
\begin{equation}
\frac{1}{\Gamma_f}\frac{d\Gamma_f}{dc_K\,d\varphi} = \frac{3}{8\pi} s_K^2 \left[1-2\kappa\,
\operatorname{Re}\!\left(z e^{2i\varphi}\right)\right] \,,
\end{equation}
with $0 \leq \varphi <2\pi$. In contrast with the $J/\psi\,\phi$ channel, 
this distribution allows us to measure the negativity $\mathcal{N} = |z|$, provided $\kappa$ is known and nonzero. We remark that cancellations in the numerator of (\ref{ec:kint}) reduce the value of $|\kappa|$, which quantifies the photon spin analysing power. A fully-differential analysis allows to keep the full photon spin information; still, simple kinematical cuts can enhance $|\kappa|$ at the cost of decreasing the signal. We assume $\kappa=0.42$ with an additional selection efficiency $f_{\rm sel}=0.38$, motivated by Ref.~\cite{Bishara:2015yta}.

Converted-photon samples in current LHCb datasets are small, due to the small branching ratio $\text{Br}(B_s^0 \to \phi \gamma) \simeq 3.4 \times 10^{-5}$~\cite{PDB}, and the photon conversion rate at the 10\% level~\cite{LHCb:2024wnx}. We therefore address sensitivity estimates for future $Z$ factories. We use CEPC as a reference. For $4.1\times10^{12}$ $Z$ bosons summed over two CEPC detectors, Ref.~\cite{Wang:2025wbn} predicts $9.43\times10^5$ $B_s^0\to\phi(\to K^+K^-)\gamma$ signal decays after cuts to suppress backgrounds.
Taking a $10\%$ conversion probability~\cite{CEPCStudyGroup:2025kmw} and assuming $30\%$ pair recovery gives approximately $3\times10^4$ signal candidates from converted photons. For initial-flavour tagging we assume $\epsilon=0.8$ at $\bar\omega=0.15$, motivated by the jet-charge tagging performance reported in Ref.~\cite{Ruan2024ICHEP}. After tagging and polarimetric selection, the nominal sample contains $9\times10^3$ events. For FCC-ee, assuming comparable selection and tagging performances and a $12\%$ conversion probability~\cite{Bacchetta:2019fmz}, rescaling to $6\times10^{12}$ $Z$ bosons~\cite{FCC:2025lpp}  (summed over four detectors) gives $1.5\times10^4$ events. For LHCb Upgrade II at the high-luminosity LHC, scaling the Run 1 yield~\cite{Beaucourt} to the projected luminosity and taking into account trigger upgrades~\cite{trigger}, gives an estimate of $2.45 \times 10^4$ converted events. An improved initial-flavour tagging could make this sample large enough for the study of entanglement oscillations. 

\begin{figure}[htb]
\begin{center}
\includegraphics[width=8cm,clip=]{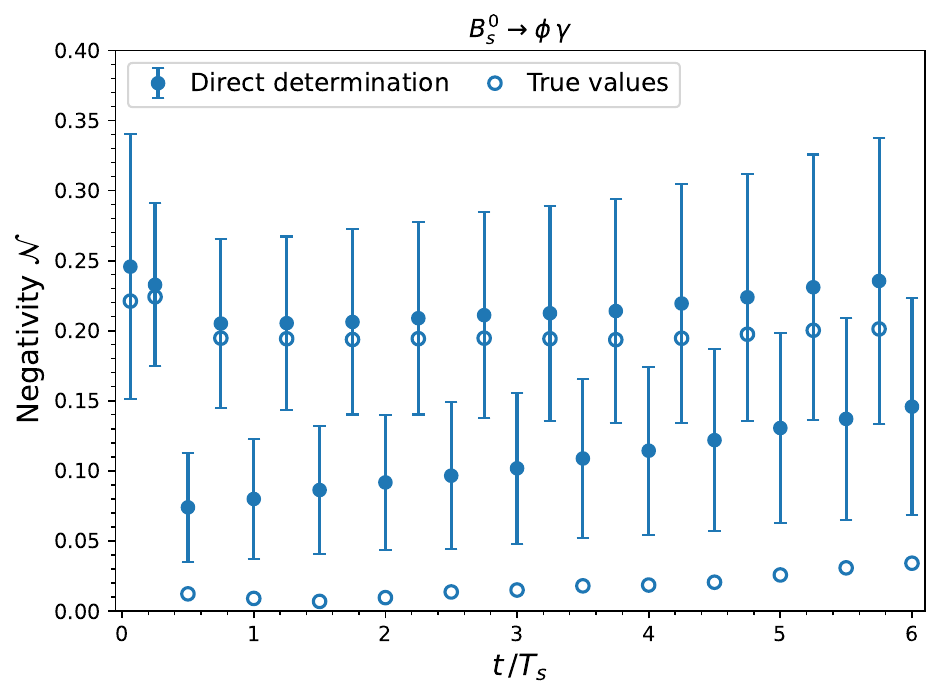}
\caption{Expected $\phi\gamma$ negativity for $1.5\times10^4$ tagged decays. Filled circles show the pseudoexperiment means with $1\sigma$ statistical errors; open circles show the true values.}
\label{fig:Ngamma}
\end{center}
\end{figure}

We estimate the sensitivity of CEPC and FCC-ee with 2000 pseudoexperiments for each collider benchmark, including mistagging and Gaussian decay-time smearing with $\sigma_t=55$~fs. This assumed resolution is motivated by anticipated improvements in vertexing and photon reconstruction
relative to the average $70$~fs achieved by LHCb in $B_s^0\to\phi\gamma$ decays~\cite{LHCb:2019vks}. As in the $J/\psi$ channel, we reverse the sign of $\varphi$ for events tagged as $\bar B_s^0$. Figure~\ref{fig:Ngamma} shows the expected negativity measurements and their $1\sigma$ statistical uncertainties for FCC-ee. One can observe that statistical fluctuations in $\hat z_i$ bias the estimated negativity $|\hat z_i|$ upwards, particularly when it is small. This bias can be accounted for by using a Rice likelihood~\cite{Rice}. Comparing the oscillating prediction with the hypothesis of constant negativity using a calibrated likelihood-ratio test, we obtain expected sensitivities of $3.1\sigma$ and $4.5\sigma$ for CEPC and FCC-ee, respectively. Details of the statistical analysis are given in the Appendix.

\section{Discussion}

In this work we have shown that the coherent flavour evolution of a neutral meson leads to an oscillating behaviour of the spin entanglement shared by its decay products. This effect exhibits the interplay between several genuinely quantum effects: (i) flavour evolution and coherence; (ii) particle decay as a projection; and (iii) spin entanglement of the decay products.
We have examined two examples of interest, $B_s^0 \to J/\psi \,\phi$ and $B_s^0 \to \phi \gamma$. A salient feature of the $J/\psi\,\phi$ channel is that the measured bounds on the negativity can certify a change in spin entanglement even without full spin tomography. In $B_s^0 \to \phi \gamma$ the connection between flavour coherence and entanglement is particularly direct. In the dominant-helicity approximation, the spin entanglement directly measures the flavour coherence of the parent meson at decay. In addition, this process provides a novel example of a quantum information test using high-energy photons, extending previous studies with fermions and weak bosons. The current LHCb sample offers a sensitivity exceeding $4\sigma$ to a
change in the $J/\psi\,\phi$ negativity, while future $Z$ factories could provide evidence of entanglement oscillations in $B_s^0 \to \phi \gamma$. Our exploratory estimates include mistagging and decay-time resolution, leaving the impact of residual backgrounds and other detector effects to a dedicated experimental study.

\section*{Acknowledgements}

JAAS thanks the CERN Theory Department for hospitality during the finalisation of this work.
This work has been supported by the Spanish Research Agency (Agencia Estatal de Investigación) through grants PID2022-142545NB-C21, PID2025-168810NB-I00, CEX2020-001007-S, and CEX2025-001574-S (research lines BSM, Quantum) funded by MCIN/AEI/10.13039/501100011033, and by the Swiss National Fund grant FNS/2000-1-243092/10009324.

\section*{Appendix: Statistical analysis}

In our statistical analysis, each pseudoexperiment represents one possible experimental outcome. For each of them, we evaluate the statistical significance for the rejection of the constant-negativity hypothesis. Each $p$-value is converted into a one-sided Gaussian significance, and quote the median significance over the pseudoexperiments as the expected sensitivity.

\subsection{$J/\psi\,\phi$ channel}

For each pseudoexperiment we evaluate the test statistic $T$, whose value we denote by $T_\text{obs}$. We then obtain a conservative bound on the probability of obtaining $T \geq T_\text{obs}$ if the negativity has a common value $N_0$ in all bine. Numbering the shortened bin at $t=0$ as bin~1, we evaluate $T$ in
bins~2--12 and use bins~13--16 as an independent reference sample. The four reference bins span one period $T_s$ and form two pairs, with the bin centres in each pair separated by $T_s/2$. In each reference bin $j$, we take the midpoint $(\widehat L_j + \widehat U_j)/2$ of the estimated negativity interval. We label the pair with the higher average midpoint as peaks and the other pair as troughs. Each earlier bin $i$ takes the label of the reference bin $j$ at the same point in the oscillation cycle, i.e.\ with centres separated by an integer multiple of $T_s$.

Second, for each peak bin $i$, we take the phases $\alpha_i=\arg\widehat w_j$ and $\beta_i=\arg\widehat z_j$ from its reference bin $j$. We use these phases with $\widehat w_i$ and $\widehat z_i$ measured in the peak bin itself to obtain $\hat L'_i$ in (\ref{ec:bounds2}). Taking the phases from bin $i$ itself would instead give $|\widehat w_i|+|\widehat z_i|$, reintroducing the upward bias from taking moduli. For a trough bin, the reference supplies only the label, since $\widehat U'_i$ requires no phase.

For each pseudoexperiment, we then seek an upper bound on the probability of obtaining $T\geq T_{\rm obs}$. Even at fixed negativity, different spin states can give different distributions of $T$, so we seek a probability bound valid for all such states. To use the Bennett~\cite{Bennett1962} and Bentkus~\cite{Bentkus2004} inequalities, for sums of independent event contributions, we construct a quantity $S$ satisfying $S \geq T$, which can be written as such a sum. In each trough bin we replace $|\widehat z_i|$ by $\RE(e^{-i\gamma_i}\widehat z_i)$, the component of $\widehat z_i$ along the direction of the true $z_i$. Since this component cannot exceed the modulus and the trough term is subtracted in (\ref{ec:TJpsi}), $S\geq T$ in every pseudoexperiment. With this choice of $\gamma_i$, the projected estimate has mean $\RE(e^{-i\gamma_i}z_i)=|z_i|$ and the mean trough contribution is $U'_i\geq N_0$.
The mean peak contribution is $L'_i\leq N_0$, so $S$ has nonpositive mean.
Since $S\geq T$ in every pseudoexperiment, every outcome with $T \geq T_\text{obs}$ also has $S \geq T_\text{obs}$. Hence the probabilities satisfy $P (T\geq T_\text{obs}) \leq P(S \geq T_\text{obs})$.
With the bin counts and reference measurements fixed, we use the Bennett and Bentkus inequalities to bound the latter probability.

Finally, in order to apply the Bennett and Bentkus inequalities, we need an upper bound on the variance of $S$. From the second moments of the angular weights we obtain an upper confidence bound on the variance of $S$, with failure probability at most $\eta=10^{-9}$. We add $\eta$ to the smaller of the Bennett and Bentkus tail bounds and set $p=1$ if the sum exceeds unity.

\subsection{$\phi\gamma$ channel}

For this channel the negativity estimator in each bin $i$ is $|\hat z_i|$, with $z$ the complex matrix element $\rho_{+-;-+}$, evaluated at that bin. We approximate the real and imaginary parts of $\hat z_i$ by independent Gaussians with common variance $s_i^2\simeq(2\kappa^2 n_i)^{-1}$, where $n_i$ is the number of events in bin $i$. In this approximation, $|\hat z_i|$ follows a Rice distribution~\cite{Rice}, with noncentrality equal to the underlying bin negativity and scale $s_i$. The likelihood is the product of these distributions over the different bins.

For each collider benchmark, the oscillating template is determined by a separate set of 2000 pseudoexperiments. In each bin we average the estimates $\hat z_i$ with weights equal to the number of events, and take the modulus of the average. The template is then kept fixed when evaluating the sensitivity. We compare it with the best-fitting constant negativity $N_0$ using a test statistic
\begin{equation}
T = 2\left[ \log\mathcal L_{\rm osc} -\max_{0\leq N_0 \leq 1/2} \log \mathcal L(N_0) \right] \,,
\end{equation}
where $\mathcal L_{\rm osc}$ and $\mathcal L(N_0)$ are the Rice likelihoods for the oscillating template and a constant negativity $N_0$, respectively. The likelihood includes bins up to $10 \, T_s$. 
Since the constant value is unknown, the $p$-value is the largest probability over allowed $N_0$ of obtaining $T$ at least as large as observed, keeping the bin counts fixed.

\end{document}